\documentclass[conference]{IEEEtran}
\usepackage{float}
\usepackage{amsmath,amsfonts,amssymb}
\usepackage{booktabs}
\usepackage{algorithm}
\usepackage{algorithmic}
\usepackage{array}
\usepackage[caption=false,font=normalsize,labelfont=sf,textfont=sf]{subfig}
\usepackage[justification=raggedright]{caption}
\usepackage{textcomp}
\usepackage{stfloats}
\usepackage{url}
\usepackage{verbatim}
\usepackage{graphicx}
\usepackage{subfig}
\usepackage[style=ieee,backend=bibtex]{biblatex}
\usepackage{wrapfig}
\usepackage{tabularx}
\usepackage{enumitem,kantlipsum}
\DeclareFieldFormat{doi}{} %de-display doi
\usepackage{cleveref}
\crefname{figure}{Fig.}{Figs.}
\Crefname{figure}{Fig.}{Figs.}
\crefname{equation}{Eq.}{Eqs.}
\Crefname{equation}{Eq.}{Eqs.}
\usepackage{xcolor}
\usepackage{soul}
\soulregister\cite7 % make ref and cite can be inclueded in comment
\soulregister\ref7
\soulregister\eqref7

\def\BibTeX{{\rm B\kern-.05em{\sc i\kern-.025em b}\kern-.08em
    T\kern-.1667em\lower.7ex\hbox{E}\kern-.125emX}}
\begin{document}
\captionsetup[subfigure]{font=small, labelfont=small} 
\renewcommand*{\bibfont}{\footnotesize}
% \DeclareFieldFormat[article]{journaltitle}{\mkbibacro{#1}}

\title{Low-Complexity Near-Field Beam Training with DFT Codebook based on Beam Pattern Analysis}
% \title{Beam Pattern Analysis of Hybrid-Field Beam Training with DFT Codebook}
\author{
\IEEEauthorblockN{Zijun Wang\IEEEauthorrefmark{1}, Rama Kiran\IEEEauthorrefmark{2},  Shawn Tsai\IEEEauthorrefmark{3}, Rui Zhang\IEEEauthorrefmark{1}}
\IEEEauthorblockA{\IEEEauthorrefmark{1}Department of Electrical Engineering, The State University of New York at Buffalo, Buffalo, NY, USA\\ Emails: {zwang267@buffalo.edu, rzhang45@buffalo.edu}}
\IEEEauthorblockA{\IEEEauthorrefmark{2}CSD, MediaTek Inc., Bengaluru, India. Email: rama.kiran@mediatek.com}
\IEEEauthorblockA{\IEEEauthorrefmark{3}CSD, MediaTek Inc., San Diego CA USA.  Email: shawn.tsai@mediatek.com}
\thanks{This paper was produced by the IEEE Publication Technology Group. They are in Piscataway, NJ.}
\thanks{Zijun Wang and Rui Zhang are with the Department of Electrical Engineering, The State University of New York at Buffalo, New York, USA (email: {zwang267@buffalo.edu}, {rzhang45@buffalo.edu}).}
}

% \IEEEpubid{0000--0000/00\$00.00~\copyright~2025 IEEE}

\maketitle
% {Backup title: One-Stage Near-Field Beam Training with DFT Codebooks: A Low-Complexity Solution}
\begin{abstract}
Extremely large antenna arrays (ELAAs) operating in high-frequency bands have spurred the development of near-field communication, driving advancements in beam training design. 
This paper introduces an efficient near-field beam training method that utilizes the discrete Fourier transform (DFT) codebook traditionally employed for far-field users (FUs). We begin by analyzing the received beam pattern and deriving closed-form expressions for its width and central beam gain as a function of user location. Building on these derived expressions, we propose a beam training method that estimates user distance from the received signal powers using DFT beams, with a computational complexity of
%This paper proposes an efficient near-field beam training method using the discrete Fourier transform (DFT) codebook that is conventionally used for far-field users (FUs). We begin by analyzing the received beam pattern and deriving a closed-form expression for its width and central beam gain as a function of user locations.  Building on this, we propose a beam training method capable of estimating user distance based on the user received signal power using DFT beam, with a complexity of 
$\mathcal{O}\left(1\right)$. Simulation results confirm the efficacy of our approach, demonstrating its ability to perform low-complexity near-field beam training with high estimation accuracy.  Notably, our proposed scheme achieves up to a 1.96 dB SNR gain over the exhaustive search methods in multi-user beamforming scenarios.

%Extremely large-scale array (XL-array) has been widely applied with the rapid development of next-generation communication systems, leading the transmission of signals to be in the near-field domain. The difference in electromagnetic (EM) propagation modeling of near-field and far-field has inspired people to create many different ways for beam training in near-field. However, in this article, we show that the DFT codebook, which is conventionally used for far-field beam forming, also have the potential for near-field beam training. We work on the beam pattern of using DFT codebook on near-field users (NUs), showing a semi-closed-form of beam width and a semi-closed-form of central beam gain. Based on the beam width, we define the modified Rayleigh distance for certain antenna elements. Then we verify the relation of beam width and beam central gain on modified Rayleigh distance. Using these relations, we raise an efficient beam training method and perform the numerical simulation. The results shows that the relation is practical in beam trainineq:G and fits well with the true beam pattern.
\end{abstract}


\section{Introduction}
%\IEEEPARstart{N}{ext}-generation communication systems demand higher spectral efficiency, which naturally leads to the adoption of extremely large-scale arrays (XL-arrays) to leverage beam forming techniques \cite{10496996}.
\IEEEPARstart{T}{he} rapid evolution of wireless communications, fueled by the growing demand for higher data rates and ultra-low latency, has spurred significant interest in terahertz bands, due to their extensive available bandwidth \cite{10438977,9766110}. The utilization of high RF bands necessitates the deployment of extremely large antenna arrays (ELAA), allowing for high data rates and enhanced coverage \cite{10496996}.  {However, the increased number of antenna elements in ELAA, coupled with the smaller wavelengths,  makes the far-field assumptions of conventional wireless communication no longer hold true for certain users. Specifically, for users located within the near-field region (i.e., tens of meters or hundreds of meters), the electromagnetic (EM) wavefront is spherical rather than planar} \cite{10220205}. Consequently, the near-field steering vector depends not only on the angle but also on the distance. Therefore, conventional beamforming and beam training methods that use a DFT-based far-field codebook to sweep the angular domain are insufficient for near-field users (NUs).  

\vspace{-1mm}
In the literature, there are proposals for beam training  codebooks that are functions of both angle and distance. For instance, \cite{9693928} proposes a polar-domain codebook for NUs, that includes angles and distances of the target NUs, which has been successfully demonstrated in beam training \cite{10239282}. However, the aforementioned codebooks exhibit high overhead, as the beam training requires exhaustive searches across both the angular and distance domains, leading to an increased codebook size. Therefore, to reduce the overhead, some novel schemes for beam training in near-field still leverage DFT codebook. \cite{10239282}  proposes a two-stage beam training procedure. In the first stage, a coarse search is performed in the angular domain using a DFT codebook on subarray. The second stage conducts a refined search across both the angular and distance domains utilizing the near-field codebook. Furthermore,  \cite{9913211}  introduces a fast two-stage beam training method, employing DFT codewords in the first stage to obtain the angle using the central portion of the beam pattern, followed by refining both the distance and angle in the second stage using near-field codebook. Similarly, \cite{10500334} demonstrates that DFT codewords can be utilized to jointly estimate the angle and distance of NUs by numerically analyzing the beam width and beam gain, yielding lower complexity than other methods. Recently, deep learning approach has also been explored to further improve beam training performance with DFT codebooks. \cite{DL_for_beam_training} proposes neural networks that uses far-field wide-beam measurements to estimate and refine the optimal near‑field beam.

\vspace{-1mm}
Despite these advancements, existing studies on near-field beam training using the DFT codebook primarily rely on numerical simulations or sophisticated artificial intelligence (AI), resulting in high computational complexity. In this paper, we propose a low-complexity near-field beam training scheme which leverages the beam pattern of beam sweeping using DFT codebook and maintains the comparable overhead with other DFT-based schemes. In particular, we first derive closed-form expressions for the received beam pattern width and gain for NUs with DFT codebook. Based on these derivations, we propose a low-complexity beam training method, which is validated through simulation analysis for both single-user and multi-user beamforming scenarios.

\vspace{-1mm}
The rest of the paper is organized as follows. \Cref{sec:system model} presents the system model. \Cref{sec:beam pattern} analyzes the received beam pattern, deriving closed-forms for beam pattern width and beam pattern gain. Based on the theoretical derivation, \Cref{sec:proposed scheme} proposed the low-complexity beam training. Simulations and comparison with benchmark schemes are presented in \Cref{sec:simulation}. \Cref{sec:clu} presents conclusion.

\section{System Model}\label{sec:system model}
\begin{figure}[!t]
\captionsetup{justification=raggedright, singlelinecheck=false} % 主图 caption 居左
\centering
\includegraphics[width=0.75\linewidth]{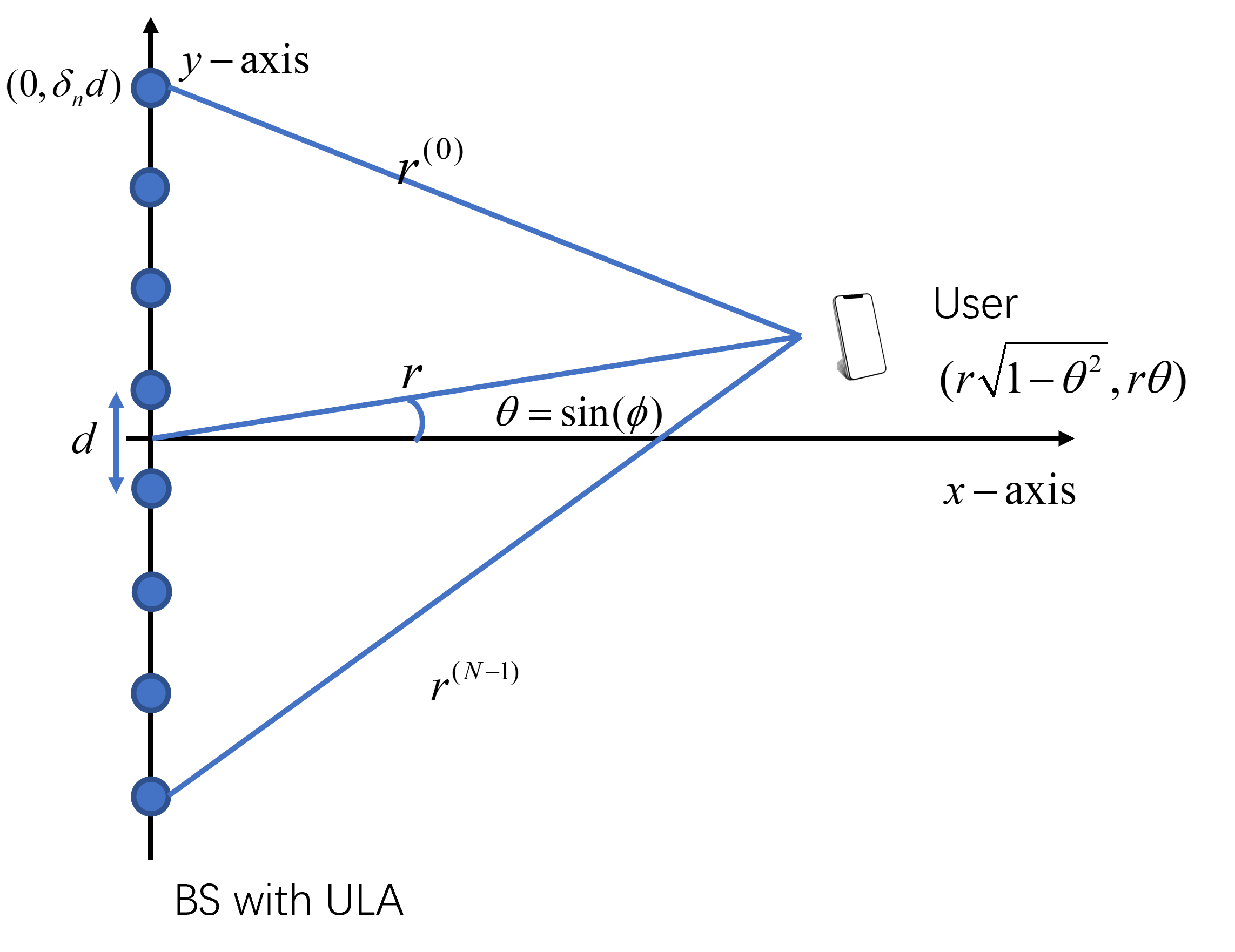}
\caption{Near-field communication system with ULA}
\label{fig:antenna}
\end{figure}

\vspace{-1mm}
Conventional method defines the near-field region as between the Fresnel distance $R_{Fre}=\frac{1}{2} \sqrt{\frac{D^{3}}{\lambda}}$ and the Rayleigh distance $R_{Ray}=\frac{2D^2}{\lambda}$, in which $\lambda$ is the wavelength at the central frequency,  $d=\frac{\lambda}{2}$ is the spacing of the antenna elements and $D\approx Nd$ is the aperture of antenna \cite{7942128}. The far field lies farther away than $R_{Ray}$. In this article, we consider a base station (BS) with the uniform linear array (ULA) consisting of $N$ antenna elements, and user is equipped with one single antenna. The BS adopts DFT codebook to perform beam training. 
As shown in \Cref{fig:antenna}, we assume that the antenna is placed on the y-axis, with the center at $(0, 0)$, and each antenna element has the coordinate of $(0,\delta_{n} d)$, with $\delta_{n}=\frac{2n-N+1}{2},n\in\mathcal{N}$$\triangleq$$ \{0,1,\dots, N-1\}$ being the index of the antenna elements.  

Since high RF bands are dominated by line-of-sight (LoS) path, we consider the following channel model \cite{9112745}.
\begin{equation}
   \mathbf{h}^{H}\left(\theta, r\right)=\sqrt{N} g e^{-\frac{\jmath 2 \pi r}{\lambda}}\mathbf{b}^{H}(\theta, r),  
   \label{eq:channel h}
\end{equation}
where $\jmath$ is $\sqrt{-1}$, $g=\frac{\lambda}{4\pi r}$ is the channel gain, and $\mathbf{b}(\theta, r)$ is the near-field steering vector of ULA that is given by
\begin{equation}
     \mathbf{b}(\theta, r)=\frac{1}{\sqrt{N}}\left[e^{-\jmath 2 \pi\left(r^{(0)}-r\right) / \lambda}, \cdots, e^{-\jmath 2 \pi\left(r^{(N-1)}-r\right) / \lambda}\right]^{T}.
     \label{eq:near-field vector}
\end{equation}
And $\theta=\sin (\phi)$ is the spatial angle of users at BS, with range $[-1,1]$. $\phi$ is the angle of departure (AoD) of the users. $r^{(n)}=\sqrt{r^{2}+\delta_{n}^{2} d^{2}-2 r \theta \delta_{n} d}$ is the physical distance from $n$-th antenna elements to users and {$r$ is the physical distance from central of the antenna array to users}. 

The far-field codeword \( \mathbf{a}(\varphi_n) \) is defined by DFT basis as follows:
\begin{equation}
\mathbf{a}(\varphi_n) \triangleq \frac{1}{\sqrt{N}} 
\left[ 
e^{-j\pi \left(-\frac{N-1}{2}\right)\varphi_n},
\cdots,e^{-j\pi \left(\frac{N-1}{2}\right)\varphi_n}
\right],\forall n \in \mathcal{N}.   
\end{equation}
For DFT codebook $\mathbf{V}_{DFT}=\{\mathbf{a}\left(\varphi_0 \right),\mathbf{a}\left(\varphi_1 \right),\cdots,\mathbf{a}\left(\varphi_{N-1} \right)\}$, we uniformly sample the angle domain such that $\varphi_{n}=\frac{2 n-N+1}{N},n\in \mathcal{N}$. Each codeword is denoted as \( \mathbf{v}_n = \mathbf{a}(\varphi_n) \).  
Based on the channel model, we denote the received signal at a user with a position $(\theta,r)$ using beam forming vector $\mathbf{v}_n$ as
\begin{equation}
  y(\mathbf{v}_n)=\mathbf{h}^{H}\left(\theta, r\right)\mathbf{v}_n s+w,  
\end{equation}
in which  %, and we ignore the noise raised by amplifier. 
$w\sim\mathcal{CN}(0,\sigma^2)$ is the complex additive white Gaussian noise (AWGN). $s$ is the reference pilot signal, and we assume that $s=1$. 
\vspace{-1mm}
\section{Received Beam Pattern Analysis}\label{sec:beam pattern}
\vspace{-1mm}
{This section analyzes the received beam pattern of using DFT codebook for beam sweeping and derives the beam pattern width and gain as a function of distance and angle. Given user location \((\theta,r)\), we consider the following beam gain using a DFT codeword on NUs}:
\begin{align}
    G\left(\mathbf{b}^{H}(\theta, r),  \mathbf{a}(\varphi) \right) &=\left|\mathbf{b}^{H}(\theta, r) \mathbf{a}(\varphi)\right|\nonumber\\
    & \approx\lvert f(\theta, r, \varphi)\rvert.
    \label{eq:G and f}
\end{align}
In the final step we use the Taylor series to approximate $r^{(n)}\approx r-\delta_{n} d \theta+\frac{\delta_{n}^{2} d^{2}\left(1-\theta^{2}\right)}{2 r}$. And we denote
\begin{align}
    &f(\theta, r, \varphi)\nonumber\\
    &=\frac{1}{N}\sum_{n=0}^{N-1} \exp \left(\jmath \pi\left[-\delta_{n}(\theta-\varphi)+\frac{\delta_{n}^{2} d\left(1-\theta^{2}\right)}{2 r}\right]\right),  
    \label{eq:f define}
\end{align}
where $\varphi \in \{\varphi_{n}\mid\varphi_{n}=\frac{2 n-N+1}{N},n\in \mathcal{N}\}$ is the angle sampled for DFT codeword and the NU location in the polar domain is denoted as \( (\theta, r) \in \mathbb{R}^2 \).
Let integral variable $x=\frac{2n-N+1}{N}\in[-1,1], n\in\mathcal{N}$, the $f(\theta, r, \varphi)$ in \Cref{eq:f define} can be further approximated as:
\begin{align} \label{eq:f_function2}
&\tilde{f}(\theta, r, \varphi) \nonumber\\
&= \frac{1}{2} \int_{-1}^{1} {\exp \left(\jmath \pi\left[\frac{N^{2} x^{2} d\left(1-\theta^{2}\right)}{8 r}-\frac{N x}{2}(\theta-\varphi)\right]\right) \operatorname{d} x}.  
\end{align}
Note that the approximation in \Cref{eq:f_function2} holds under the conditions $(\theta-\varphi)\ll2,\frac{Nd}{r}\ll2$ according to the Riemann Integral.
Then, we denote $\alpha=\frac{N^{2} d\left(1-\theta^{2}\right)}{8 r}$ and $\beta=\frac{N(\theta-\varphi)}{2}$, and substitute them into \Cref{eq:f_function2},  the function \( \tilde{f}(\theta, r, \varphi) \) becomes
   \begin{align}
       & \tilde{f}(\theta, r, \varphi) \nonumber\\
   &=\frac{e^{\jmath\left(\frac{-\beta^{2}+\alpha}{4 \alpha}\right)\pi\Bigl[\operatorname{erf}\left(\frac{e^{\jmath\frac{3}{4} \pi}\sqrt{\pi}(\beta-2 \alpha) }{2 \sqrt{\alpha}}\right)-\operatorname{erf}\left(\frac{e^{\jmath \frac{3}{4} \pi}\sqrt{\pi}(\beta+2 \alpha) }{2 \sqrt{\alpha}}\right)\Bigr]}}{ 4\sqrt{\alpha}}.   
   \end{align}
    \label{eq:origin f} 
% {
% \begin{equation}
%    \begin{split}
%     & \tilde{f}(\theta, r, \varphi) \\
%    &=\frac{e^{\jmath\left(\frac{-\beta^{2}+\alpha\pi}{4 \alpha}\right)}}{ 4\sqrt{\alpha}}\Bigl[\operatorname{erf}\left(\frac{e^{\jmath\frac{3}{4} \pi}\sqrt{\pi}(\beta-2 \alpha) }{2 \sqrt{\alpha}}\right)\\
%    &\quad\quad\quad\quad\quad\quad\quad\quad\quad-\operatorname{erf}\left(\frac{e^{\jmath \frac{3}{4} \pi}\sqrt{\pi}(\beta+2 \alpha) }{2 \sqrt{\alpha}}\right)\Bigr].
%     \label{eq:origin f} 
% \end{split} 
% \end{equation}
% }
where the error function is defined as $\operatorname{erf}(x)=\frac{2}{\sqrt{\pi}} \int_{0}^{x} e^{-y^{2}} \operatorname{d} y$. The detailed calculation can be found in Appendix.
% The detailed calculation can be found in \Cref{app:close-form}.
% \hlc{appendix A or Appenidx?}
In the following, we first derive the central beam pattern gain, and then use it to derive the beam width.

\subsection{Derivation of Central Beam Pattern Gain}\label{subsec:Central Beam Pattern Gain}
We define the central beam pattern gain as the value of $G\left(\mathbf{b}^{H}(\theta, r), \mathbf{a}(\varphi)\right)$ when $\varphi=\theta  $ (i.e., $\beta=0$). The \Cref{eq:origin f} can be re-written to 
\begin{equation}
    \tilde{f}(\theta, r, \theta)= - \frac{1}{2}\frac{e^{\jmath\frac{\pi}{4}} \, \mathrm{erf}\left[e^{\jmath\frac{3}{4}\pi} \sqrt{\alpha} \sqrt{\pi}\right]}{\sqrt{\alpha}}.
\label{eq:f function approx}
\end{equation}
The XL-array consists of a large number of antenna elements, making \( N \) significantly large. Using the Taylor series expansion, when \(\alpha = \frac{N^2 d (1 - \theta^2)}{8r} \to \infty\), the function can be approximated as
\begin{equation}
   2 \tilde{f}(\theta, r, \theta) = (-1)^{\frac{1}{4}} \sqrt{\frac{1}{\alpha}} + O\left(\frac{1}{\alpha}\right)^{\frac{3}{2}} .
\label{eq:f_theta_r_phi}
\end{equation}
Here, \((-1)^{\frac{1}{4}} \sqrt{\frac{1}{\alpha}}\), represents the leading contribution to the amplitude of \(f(\theta, r, \varphi)\), and thus the central beam gain can be expressed as
\begin{equation}
G\left(\mathbf{b}^H(\theta, r), \mathbf{a}(\theta)\right) \approx \frac{1}{2\sqrt{\alpha}},
\label{eq:G_theta_r}
\end{equation}
% However, when \(r\) becomes excessively large, \(\alpha\) can no longer be considered as approaching infinity. In such cases, the Taylor series approximation used in the derivation fails and the above equation is no longer valid.
% \hlc{We can further write the central beam gain by accounting for the path loss as:} According to the channel assumption, the amplitude of beam gain will decrease with distance $r$. We define
% \begin{equation}
%     G_2\left(\mathbf{b}^{H}(\theta, r), \mathbf{a}(\varphi)\right) =|\mathbf{h}^{H}\left(\theta, r\right)\mathbf{a}(\varphi)|,
% \end{equation}
% \hlc{deleate eq(14), it is redundant}
% and therefore the central beam gain of $G_2\left(\mathbf{b}^{H}(\theta, r),\mathbf{a}(\varphi)\right)$ satisfies
% \begin{equation}
%     G_2\left(\mathbf{b}^{H}(\theta, r),\mathbf{a}(\theta)\right)\propto\sqrt{\frac{1}{r^2\alpha}}=\sqrt{\frac{8}{N^2dr(1-\theta^2)}}.
%     \label{eq:closed-form of beam gain}
% \end{equation}
\subsection{Derivation of Beam Pattern Width}\label{subsec:Analysis of Beam Pattern Width}

We define the normalized beam pattern gain as follows by normalizing the beam pattern using the central beam pattern gain defined in \Cref{subsec:Central Beam Pattern Gain}
\begin{equation}
G_{\text{norm}}\left(\mathbf{h}^H(\theta, r), \mathbf{a}(\varphi)\right)=\frac{\lvert \mathbf{h}^H(\theta,r)\mathbf{a}(\varphi)\rvert}{\lvert \mathbf{h}^H(\theta,r)\mathbf{a}(\theta)\rvert}
    \label{eq:normalized G_2}
\end{equation}
Following the approach in \cite{10185619}, we define the beam pattern width as the angular range of $\varphi$ values for which the value of $G_{\text{norm}}\left(\mathbf{b}^{H}(\theta, r),  \mathbf{a}(\varphi) \right) $ is relatively big with certain user at a position of $(\theta, r)$. To be specific, we define following beam pattern width
\begin{equation}
    \Phi = \{\varphi \mid G_{\text{norm}}\left(\mathbf{h}^{H}(\theta, r),\mathbf{a}(\varphi)\right) > \rho\},
    \label{Main angle set for G}
\end{equation}
\begin{equation}
    B(\theta,r) = \operatorname{Range}(\Phi) ,
    \label{eq:beam_width_definition}
\end{equation}
where $\Phi$ is the set of corresponding $\varphi$ at which $G_{\text{norm}}\left(\mathbf{h}^{H}(\theta, r),\mathbf{a}(\varphi)\right)$ has larger value than threshold $\rho$ and $\rho$ is within moderate $\beta$ region. We call the angle set like this \textbf{Main Angle Set}. $\operatorname{Range}(\Phi)=\operatorname{max}(\Phi)-\operatorname{min}(\Phi)$ is the range of $\Phi$.

\begin{figure}[!t]
\captionsetup{justification=centering, singlelinecheck=true} % caption 居中
\centering
\includegraphics[width=2.5in]{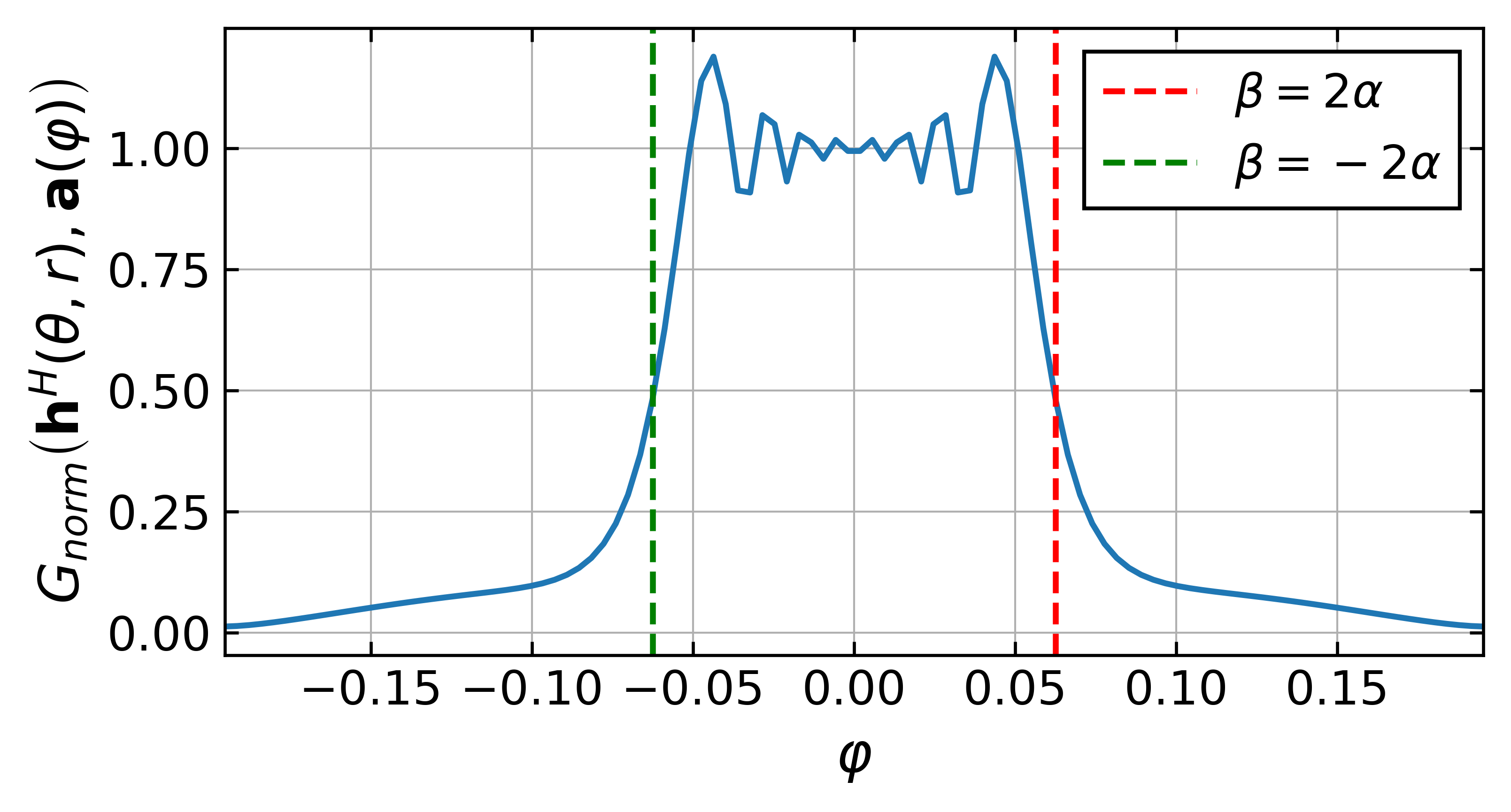}%
\captionsetup{justification=justified, singlelinecheck=false} % 主图 caption 居左
\caption{The normalized near-field beam pattern using DFT codebook with the cutting threshold $\rho=1/2$.}
\label{fig_solution}
\end{figure}

According to \Cref{eq:channel h,eq:G_theta_r,eq:origin f,eq:G and f}, we rewrite \Cref{eq:normalized G_2} as
\begin{equation}
\begin{split}
&G_{\text{norm}}\left(\mathbf{h}^H(\theta, r), \mathbf{a}(\varphi)\right)\\
&=\frac{ G\left(\mathbf{b}^H(\theta,r),\mathbf{a}(\varphi)\right)}{G\left(\mathbf{b}^H(\theta,r),\mathbf{a}(\theta)\right)}\\
&=\frac{1}{2} \left|\operatorname{erf}\left(\frac{e^{\jmath\frac{3}{4} \pi}\sqrt{\pi}(\beta-2 \alpha) }{2 \sqrt{\alpha}}\right)-\operatorname{erf}\left(\frac{e^{\jmath \frac{3}{4} \pi}\sqrt{\pi}(\beta+2 \alpha) }{2 \sqrt{\alpha}}\right)\right|.
\end{split}
\label{eq:normalized G_final}
\end{equation}
When $\beta>0$, as $\alpha$ is large, one can have $\operatorname{erf}\left(\frac{e^{\jmath\frac{3}{4} \pi}\sqrt{\pi}(\beta+2 \alpha) }{2 \sqrt{\alpha}}\right)\to -1$. For any $\rho$, we can solve $\beta$ by following equation
\begin{equation}
    \rho=\frac{1}{2} \left|\operatorname{erf}\left(\frac{e^{\jmath\frac{3}{4} \pi}\sqrt{\pi}(\beta-2 \alpha) }{2 \sqrt{\alpha}}\right)+1\right|.
    \label{eq:solve_rho}
\end{equation}
By taking $\rho=\frac{1}{2}$, one can have
\begin{equation}
    \beta-2\alpha=0.
    \label{eq:equation}
\end{equation}
%This point on the normalized beam gain is unique and the normalized beam gain is soomth within this range of $\varphi$,
Due to symmetric of the $G_{\text{norm}}\left(\mathbf{h}^{H}(\theta, r),\mathbf{a}(\varphi)\right)$ w.r.t. $\beta$, when $\beta<0$, similar conclusion can be derived and we have $\beta+2\alpha=0$. \Cref{fig_solution} illustartes an example of the above conclusions 
%when $\alpha=8$ and map $\beta=\frac{N(\theta-\varphi)}{2}$ to $\varphi$. The simulation setting is
when $N=512$ , carrier frequency $f_c=100\text{GHz}$ and {$\theta=0, r=8\text{m}$}.
%$d=\frac{\lambda}{2}=1.5$mm. 
It can be seen that the solution at either side is unique, which locates in the smooth region of the normalized beam pattern.

From $\alpha=\frac{N^{2} d\left(1-\theta^{2}\right)}{8 r}$ and $\beta=\frac{N(\theta-\varphi)}{2}$, one can obtain the beam pattern width as a function of $\theta$ and $r$:
\begin{equation}
    B(\theta,r)=\frac{Nd(1-\theta^2)}{r}.
    \label{eq:close_form of B}
\end{equation}
This is the closed-form beam width when threshold $\rho$ in \Cref{Main angle set for G} is $\frac{1}{2}$. We will use \Cref{eq:close_form of B} to guide our following beam training design. 

\section{Proposed Scheme}\label{sec:proposed scheme}

In this section, we propose a near-field beam training scheme based on the derived beam width in previous section.
\subsection{Estimate Angle by Median angle}\label{subsec:Estimate Angle by median angle}

For a user located at  $(\theta,r)$, we first perform beam sweeping using the DFT codebook to obtain the receive signal amplitude $|y(\mathbf{v}_n)|$
\begin{equation}
    \{|y(\mathbf{v}_n)|\mid y(\mathbf{v}_n)=\mathbf{h}^{H}\left(\theta, r\right)\mathbf{v}_n x+w ,\mathbf{v}_n\in\mathbf{V_{DFT}}\}.
\end{equation}

We adopt and improve the method in \cite{9913211}, which estimates the angle of users $\theta$ by the median angle set of the beam pattern. 
%$\{|y(\mathbf{v_n})|\}$. %The details are as follows
% \begin{equation}
%     \Phi^{\rho_2} = \{\varphi_n \mid |y(\mathbf{v}_n)| > \rho_2\},
% \end{equation}
% \begin{equation}
%     \hat{\theta}=\operatorname{Med}(\Phi^{\rho_2}),
% \end{equation}
To enhance the median angle method, especially under low SNR conditions or in the presence of interference, we incorporate a clustering step that groups strong beam pattern gains.
Define the set of candidate indices as
\begin{equation}
\mathcal{I} = \{ n \mid |y(\mathbf{v}_n)| > \rho_2 \}.
\end{equation}
Next, we partition \(\mathcal{I}\) into clusters based on proximity. Specifically, two consecutive indices \(n\) and \(n'\) belong to the same cluster if \(n' - n \le L\),where \(L\) is a fixed small number. 
If \(L\) is too large, peaks due to noise variations may be grouped with the main-beam peaks, contaminating the cluster and degrading the angle estimation. Conversely, if \(L\) is too small, adjacent peaks  may be split into separate clusters. In our algorithm, we choose \(L=8\).

After grouping the indices into clusters, we examine each cluster by finding the strongest received signal within it. The cluster with the strongest received signal power is selected as the optimal cluster $C^*$. 
The main angle set is 
\begin{equation}
\Phi^{\rho_2} = \{\varphi_n \mid |y(\mathbf{v}_n)|>\rho_2,n \in C^*\}.
\end{equation}
In our implementation, the threshold $\rho_2$ is chosen as 65\% of the maximum received signal amplitude, i.e., $\rho_2=0.65\max{|y(\mathbf{v}_n)|}$. This is a different value from \cite{9913211}, however, this threshold will bring increase in performance at low SNR scenario, and thus we use $\rho_2$ for fair comparison. Finally, we estimate the user angle as the median angle of \(\Phi^{\rho_2}\)
\begin{equation}
\hat{\theta} = \frac{\max(\Phi^{\rho_2})+\min(\Phi^{\rho_2})}{2}.
\label{estimate_theta}
\end{equation}
where the estimation of $\theta$ is denoted as $\hat{\theta}$. A median-k-selection scheme is also proposed in \cite{9913211}, which choose k angles near the $\hat{\theta}$ in $\Phi^{\rho_2}$. 
\begin{algorithm}[H]
\caption{Proposed Near-Field Beam Training Scheme}
\label{alg:proposed}
\begin{algorithmic}[1]
\REQUIRE DFT codebook $\mathbf{V}_{\mathrm{DFT}}$, parameters $(N,d,\Delta_\theta,L,k)$.
\ENSURE Estimated angle $\hat{\theta}^*$, distance $\hat{r}^*$, and beamforming codeword.
\STATE \textbf{Angle Estimation:}
\begin{enumerate}
    \item Perform beam sweeping to obtain $\{|y(\mathbf{v}_n)|\},\rho_2$.
    \item Cluster indices and select the cluster $C^*$ with the highest receive signal amplitude.
    \item Form Main Angle Set $\Phi^{\rho_2}=\{\varphi_n:|y(\mathbf{v}_n)|>\rho_2, n\in C^*\}$.
    \item Estimate $\hat{\theta}$ as the median angle from $\Phi^{\rho_2}$ and select $k$ candidate angles near $\hat{\theta}$, denoted as $\hat{\theta}_1,\hat{\theta}_2,\dots,\hat{\theta}_k$.
\end{enumerate}
\STATE \textbf{Distance Estimation:}
\begin{enumerate}
    \item \textbf{For} $i=1$ \textbf{to} $k$ \textbf{do:}
    \begin{enumerate}
        \item Normalize the beam gains for candidate $\hat{\theta}_i$: compute $\dfrac{y(\mathbf{v}_n)}{y(\mathbf{a}(\hat{\theta}_i))}$.
        \item Find Main Angle Set $\hat{\Phi}_i=\Big\{\varphi_n:\Big|\dfrac{y(\mathbf{v}_n)}{y(\mathbf{a}(\hat{\theta}_i))}\Big|>\frac{1}{2}\Big\}$ and $\hat{B}_i=\operatorname{Range}(\hat{\Phi}_i)$.
        \item Solve for $\hat{r}_i$ via Eq.~(\ref{eq:estimated_r_rho}) and compute the corresponding codeword $\mathbf{b}(\hat{\theta}_i,\hat{r}_i)$.
    \end{enumerate}
     End \textbf{for}.
    \item Perform beamforming using $k$ $\mathbf{b}(\hat{\theta}_i,\hat{r}_i)$ and select the codeword with the highest received power.
\end{enumerate}
\RETURN $(\hat{\theta}^*,\hat{r}^*)$ and $\mathbf{b}(\hat{\theta}^*,\hat{r}^*)$.
\end{algorithmic}
\end{algorithm}
\subsection{Estimate Distance by Beam Width}\label{subsec:estimate_distance}
With the estimated angle $\hat{\theta}$ in Section IV.B, the beam width of the received signal is defined as follows
\begin{equation}
\begin{split}
   & \hat{\Phi} = \left\{\varphi_n \mid \left|\frac{y(\mathbf{v}_n)}{y(\mathbf{a}(\hat{\theta}))}\right| > \frac{1}{2}\right\},\\
    &\hat{B} = \operatorname{Range}(\hat{\Phi}) .
\end{split}
\label{eq:beam width}
\end{equation}
%Here $\mathbf{a}(\hat{\theta})$ is the sampled DFT codeword using the estimated user angle,
where ${y(\mathbf{a}(\hat{\theta}))}$ and $y(\mathbf{v}_n)$ are the received signals when BS is using codewords $\mathbf{a}(\hat{\theta})$ and $\mathbf{v}_n$, respectively. And $\hat{B}$ is the estimated beam width for user $(\theta,r)$. By solving \Cref{eq:close_form of B} the distance of this user is estimated as
\begin{equation}
\hat{r}
= \frac{d\bigl(1-\hat{\theta}^2\bigr)}{\bigl(\hat{B}\bigr)^2}.
\label{eq:estimated_r_rho}
\end{equation}
Then near-field codeword $\mathbf{b}(\hat{\theta},\hat{r})$ in \Cref{eq:near-field vector} is adopted. If a median-k-selection is used for estimation of $\theta$, we will find the estimated $r_i$ for all k $\hat{\theta_i}$, where $i=1,2,...k$. Then BS performs additional beam training using k near-field codewords $\mathbf{b}(\hat{\theta_i},\hat{r_i})$. The codeword with the highest received power will finally be adopted. The detailed algorithm is in \Cref{alg:proposed}.

\section{Numerical Simulation}\label{sec:simulation}

This section presents the simulation results to validate our derivations and the proposed beam training scheme. The simulation setting is the same as \Cref{subsec:Analysis of Beam Pattern Width}. We set the reference SNR {as the SNR of a user at $(0, 5\text{m})$ without beamforming}. With fixed transmitted power, {we sweep different noise power by setting reference SNR from 4 dB to 30 dB. We set the user locations within the Rayleigh distance and thus all users are NUs.} 

\begin{figure}[!t]
\captionsetup{justification=raggedright, singlelinecheck=false} % 主图 caption 居左
\centering
\includegraphics[width=2.5in]{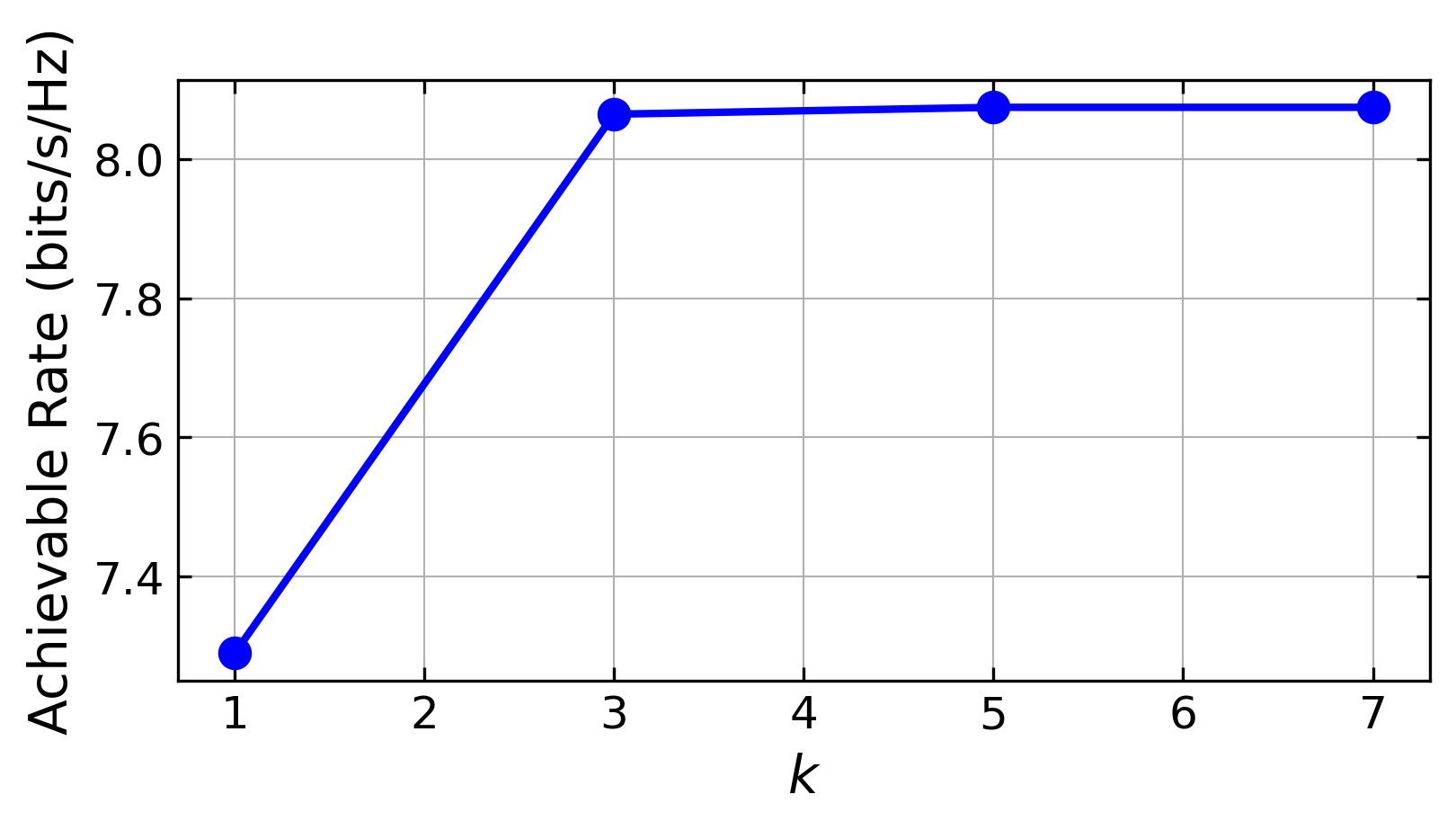}
\caption{Achievable rate versus k}
\label{fig:k_achievable_rate}
\end{figure}

Firstly, we test the influence of different k using the median-k-selection scheme on the proposed approach. In a single-user scenario, the BS utilizes the beamforming vector $\mathbf{v}=\mathbf{b}(\hat{\theta}^*,\hat{r}^*)$, which is determined using the estimated angle $\hat{\theta}^*$ and distance $\hat{r}^*$.  The achievable rate is defined as $R=\operatorname{log_2}(1+\frac{|\mathbf{h}^{H}\left(\theta, r\right)\mathbf{v}|^2}{\sigma^2})$. The SNR is set at $6$dB and the result can be found in \Cref{fig:k_achievable_rate}. It can be seen that $k=3$ has a much better achievable rate than $k=1$. When $k>3$, the improvement of achievable rate by increasing $k$ is minor. Thus for the sake of simplicity, we use $k=3$ for simulation in the rest of our analysis.

\begin{figure}[!t]
\captionsetup{justification=raggedright, singlelinecheck=false} % 组图 caption 居左
\centering
\subfloat[\tiny(a)][\textrm{\small NMSE of angle versus reference SNR.}]{%
    \includegraphics[width=0.5\linewidth]{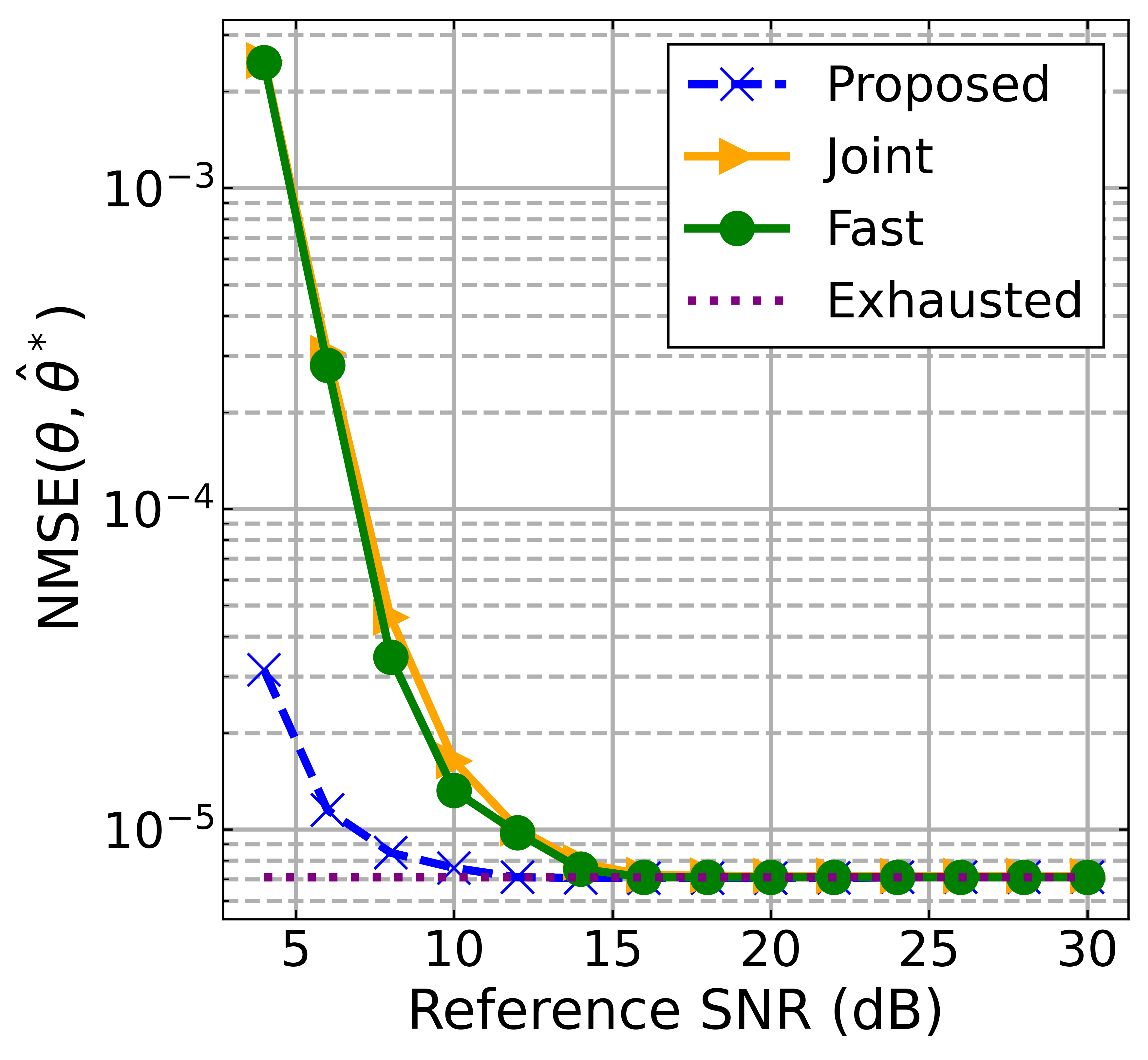}%
    \label{Angle NMSE}%
}%
% \hfill
\subfloat[\tiny(a)][\textrm{\small NMSE of distance versus reference SNR.}]{%
    \includegraphics[width=0.5\linewidth]{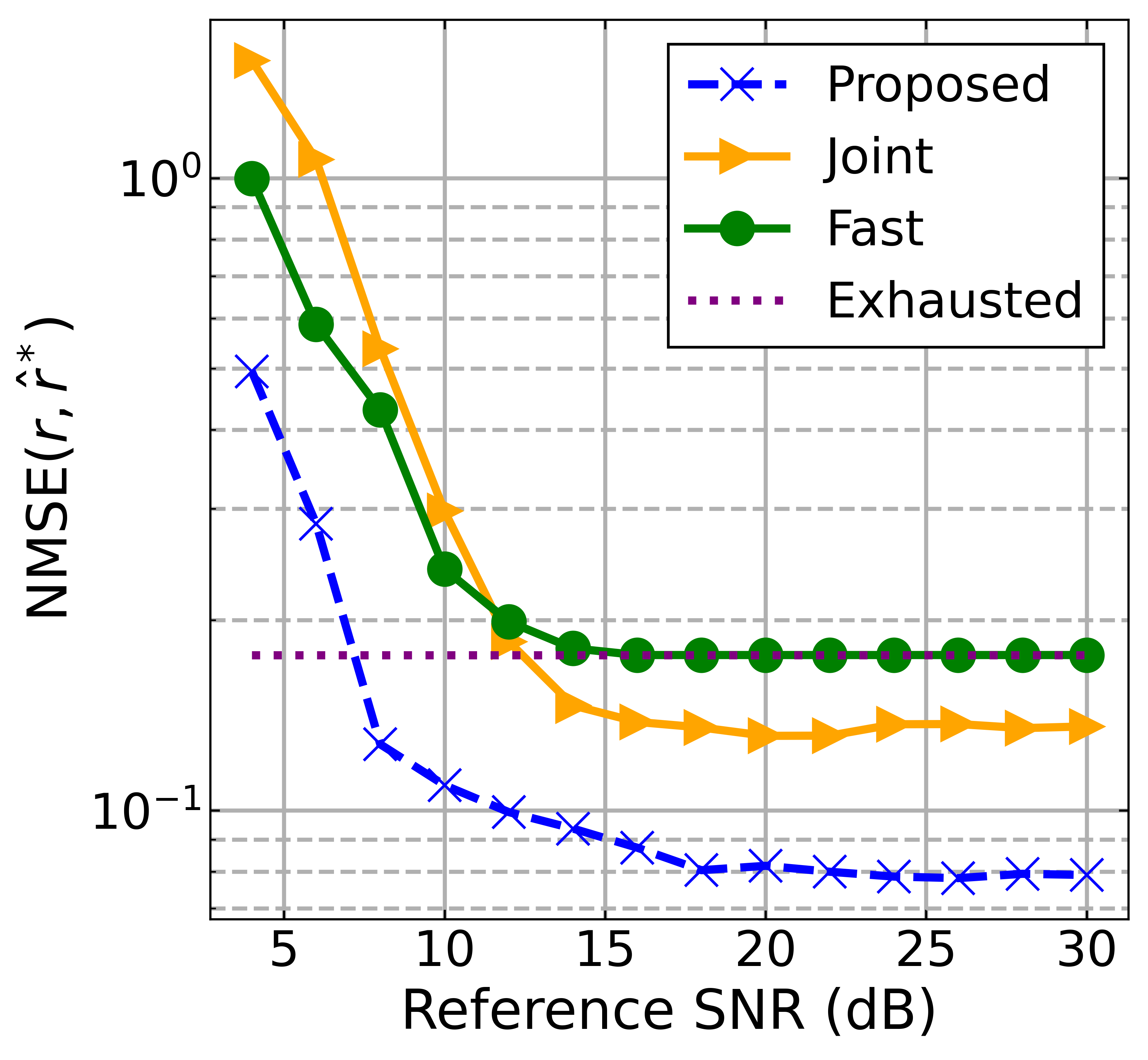}%
    \label{Distance NMSE}%
}
\captionsetup{justification=justified, singlelinecheck=false} % 组图 caption 居左
\caption{NMSE performance of (a) angle and (b) distance estimation versus reference SNR.}
\label{fig:NMSEperformance}
\end{figure}

Next we evaluate the accuracy of proposed schemes. For the true user angles \(\theta\) and their estimations \(\hat{\theta}^*\), the normalized mean square error (NMSE) is defined as  
\(\operatorname{NMSE}(\theta,\hat{\theta}^*) = \frac{\mathbb{E}\bigl[(\theta - \hat{\theta}^*)^2\bigr]}{\operatorname{Var}(\theta)}\).  
Similarly, for the true distances \(r\) and their estimations \(\hat{r}^*\), the NMSE is given by  
\(\operatorname{NMSE}(r,\hat{r}^*) = \frac{\mathbb{E}\bigl[(r - \hat{r}^*)^2\bigr]}{\operatorname{Var}(r)}\).
We compare our approach with the following schemes:
\begin{enumerate}
\item{Joint angle and distance estimation (joint scheme).} This is proposed  in \cite{10500334}.  It first uses median-k-selection scheme without clustering trick to estimate the angle, then uses approximated angle support to estimate the distance. We set  $k=3$ for a fair comparison.
\item{Fast estimation using DFT code (fast scheme).} This is proposed in \cite{9913211}. It first uses the same method as joint scheme to estimate the angle, then performs exhaustive search in distance domain to estimate distance. We also consider $k=3$.
\item{Exhausted beam training (exhausted scheme).} This is to use the near-field codebook proposed in \cite{9693928} to perform exhaustive search in both angular domain and distance domain. We set $\beta=1.6$, which is the same as in \cite{9693928}.
\end{enumerate} 
%For our proposed scheme, we also consider $k=3$,
\Cref{Angle NMSE,Distance NMSE} show the estimation accuracy of different schemes versus SNR. Our proposed scheme achieves the lowest NMSE in both angle and distance estimation. For angle estimation, as seen in \Cref{Angle NMSE}, the clustering method enables lower NMSE than other schemes in low SNR regions. For distance estimation as shown in \Cref{Distance NMSE}, the proposed scheme consistently outperforms the others, especially at higher SNR levels, with NMSE improvements of up to one order of magnitude.

\begin{figure}[!t]
\captionsetup{justification=justified, singlelinecheck=false} % 主图 caption 居左
\centering
\includegraphics[width=3.0in]{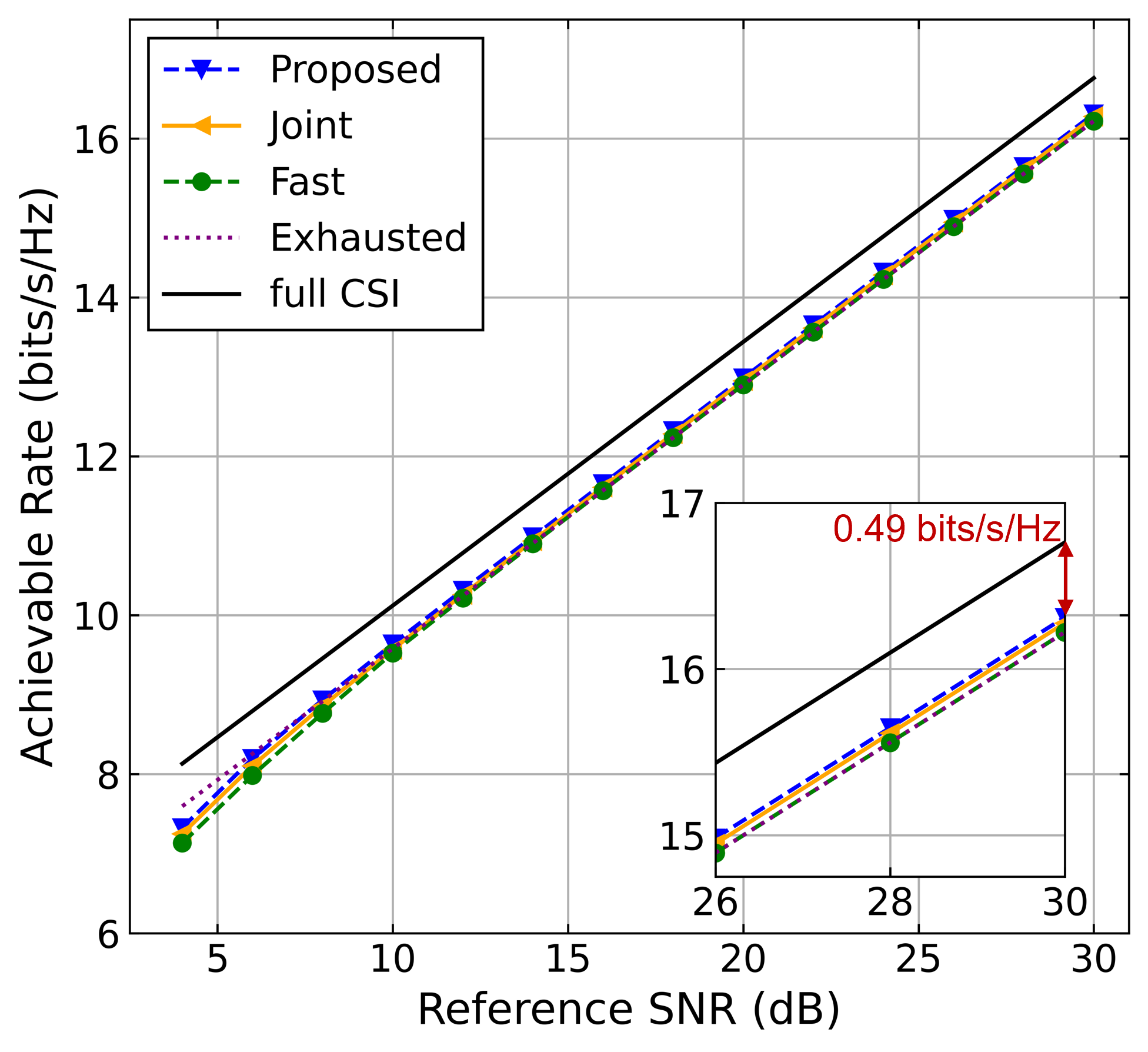}
\caption{Achievable rate versus reference SNR of single-user beamforming}
\label{Achievable rate versus reference SNR}
\end{figure}

\begin{figure}[!t]
\captionsetup{justification=justified, singlelinecheck=false} % 主图 caption 居左
\centering
\includegraphics[width=3.0in]{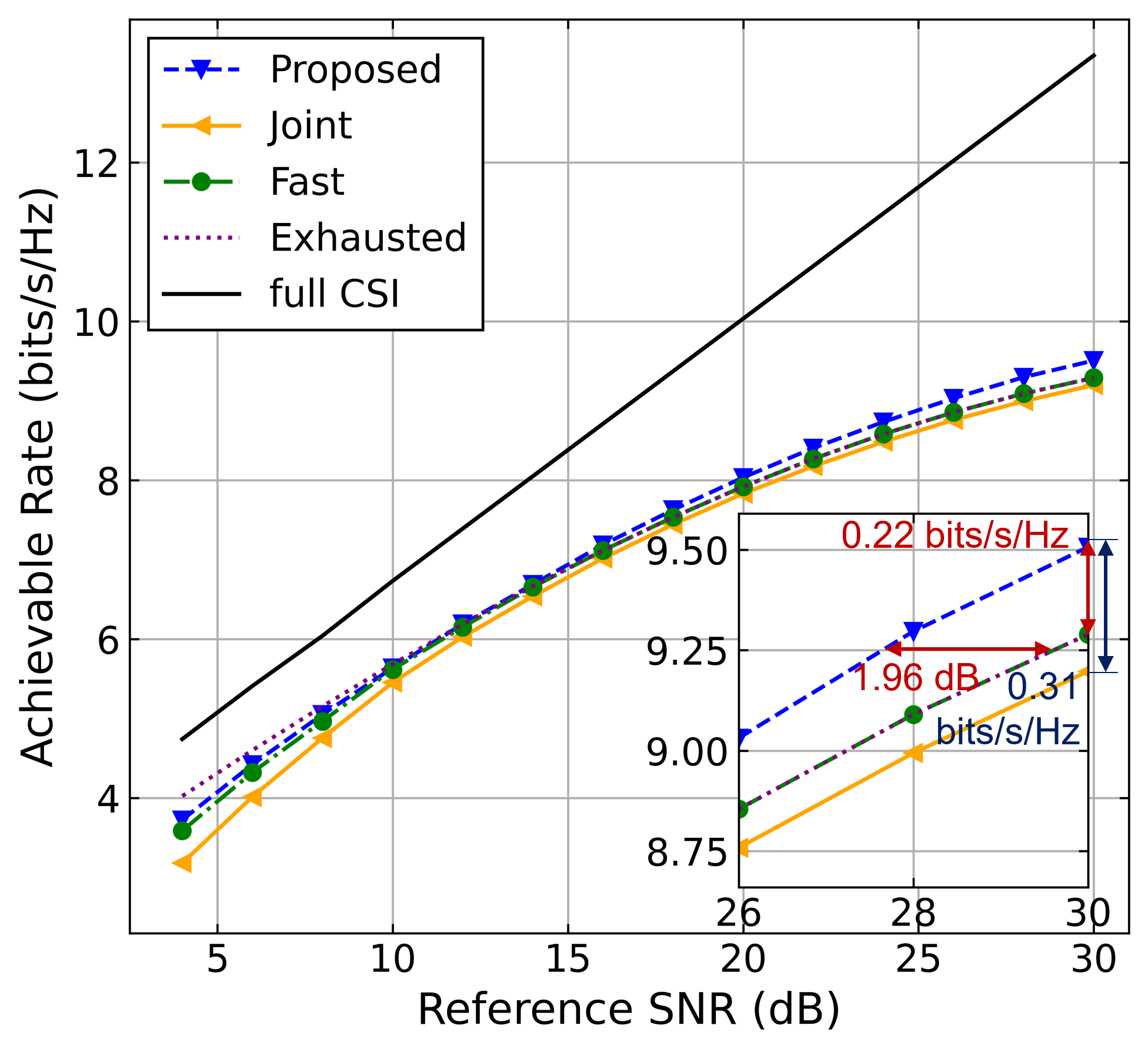}
\caption{Achievable rate versus reference SNR of multi-user beamforming}
\label{Achievable rate versus reference SNR2}
\end{figure}

Then, we evaluate the achievable rate using different beam training schemes. We begin with the single-user scenario. We also consider full channel information (CSI) as a baseline, for which the true position $(\theta,r)$ is known and $\mathbf{v}=\mathbf{b}(\theta,r)$ is set as beamforming codeword. The result is shown in \Cref{Achievable rate versus reference SNR}. Overall, the proposed method exhibits performance comparable to other schemes, with a gap of up to $0.49 \text{ bits/s/Hz}$ relative to full CSI. Our scheme achieves the highest achievable rate in high-SNR scenarios, offering an improvement of up to $0.09\text{ bits/s/Hz}$ compared with the exhausted scheme.
%\hlc{add more details here. gap between full CSI.}
Next, we analyze the performance of a multi-user beamforming scenario. We randomly group \(M = 10\) users within near-field region and perform digital beamforming following the method in \cite{9738442}, thereby serving all \(M\) users within a single time slot.  The average achievable rate is utilized for performance evaluation. For each user \(u\) with position \((\theta_u, r_u)\) (and corresponding channel \(\mathbf{h}(\theta_u, r_u)\)), the achievable rate is computed as
\begin{equation}
R_u = \log_2\!\Biggl(1+\frac{\bigl|\mathbf{h}^H(\theta_u, r_u)\,\mathbf{v}_u\bigr|^2}{\sum_{s\neq u}\bigl|\mathbf{h}^H(\theta_u, r_u)\,\mathbf{v}_s\bigr|^2 + \sigma^2}\Biggr),
\end{equation}
where \(\mathbf{v}_u\) is the beamforming vector allocated to the user at position \((\theta_u, r_u)\), \(\sigma^2\) denotes the noise power, and the denominator represents the interference from other users. For full CSI scheme, the true position \((\theta, r)\) is known and thus the beamforming codeword is optimized following the method in \cite{9738442} with true position. As shown in \Cref{Achievable rate versus reference SNR}, the proposed method achieves the best performance at high SNR value, with improvements of up to $0.22 \text{ bits/s/Hz}$ over the fast and exhausted schemes, and $0.31$ bits/s/Hz compared to the joint scheme. The gap between full CSI and our proposed scheme arises from its heavy reliance on precise channel information in multi-user beamforming.
Small errors in angle and distance estimation can lead to significant rate loss, especially when SNR is high. On the other hand, our method is more precise than other schemes as demonstrated in \Cref{fig:NMSEperformance}, which leads to the improvement in the achievable rate in this case. For instance, at an achievable rate ($9.25$ bits/s/Hz), our proposed scheme achieves the SNR gain of 1.96 dB than the fast and exhaustive schemes, as shown in \Cref{Achievable rate versus reference SNR2}.

Finally, we compare the beam sweeping overhead and complexity. The results can be seen in \Cref{table:overhead}, where $k$ median angles are selected, $S>1$ is the average sampled distance at different angle in the near-field codebook, and  $N$ is the number of angles sampled in the angular domain, which is set the same for DFT and near-field codebook for fair comparison. The overhead of our scheme is the lowest since we only samples in angular domain for beam training. 
Then we compare the complexity of different schemes 
 as shown in \Cref{table:overhead}. The 
 exhausted scheme has one single stage and the other schemes have two stages. 
 Although the joint scheme and proposed scheme have the same sweeping overhead, the proposed scheme has the lowest complexity. %Although we introduce a clustering method, 
The complexity of proposed scheme in the angle estimation stage is the same as that of the joint and fast scheme, which is $\mathcal{O}(N)$. 
%as we finalize the clustering and peak selection process at the same loop. 
%The main difference of complexity is raised by distance estimation stage.
However, our proposed scheme exhibits a lower complexity in the distance estimation stage. According to \cite{10500334}, the complexity of joint scheme is $\mathcal{O}\left(\left|\mathcal{Z}_{\mu}\right| N\right)$, where $\mathcal{Z}_{\mu}>1$ is a factor for the search by distance. However the proposed coarse method doesn't need to search and can directly calculate the distance. Therefore, the complexity is $\mathcal{O}\left(1\right)$. 

\vspace{2mm}
\begin{table}[h]
\captionsetup{justification=justified, singlelinecheck=false} % 主图 caption 居左
\caption{Training Overhead and Complexity for Different Beam Training Schemes}
\label{table:overhead}
\centering
\begin{tabular}{l c c}
\toprule
Scheme     & Training Overhead & Complexity\\
\midrule
Joint      & \(N+k\)           & \(\mathcal{O}(N)+\mathcal{O}\left(\left|\mathcal{Z}_{\mu}\right|N\right)\)\\
Proposed   & \(N+k\)           & \(\mathcal{O}(N)+\mathcal{O}(1)\)\\
Fast       & \(N+k\cdot S\)  & \(\mathcal{O}(N)\)+\(\mathcal{O}(k\cdot S)\) \\
Exhausted  & \(N\cdot S\)       & \(\mathcal{O}(N\cdot S)\) \\
\bottomrule
\end{tabular}
\end{table}

\section{Conclusion}\label{sec:clu}
In conclusion, this paper analyzes the received beam patterns of NUs using a DFT codebook and introduces a cost-effective beam training method that leverages the inherent properties of the beam pattern. We derived closed-form expressions for both the beam pattern width and the central gain, enabling efficient and accurate near-field beam training. Our proposed scheme not only exhibits low overhead and complexity but also delivers excellent performance, particularly in high-SNR scenarios.
In the multi-user beamforming case,  the proposed method achieves improvements of up to $0.31$ bits/s/Hz in achievable rate and a 1.96 dB SNR gain compared with existing schemes.

\appendix
\section*{Derivation of the Closed-Form Expression for \(f(\theta,r,\varphi)\)}\label{app:close-form}
{\small
We start from
\begin{equation}
I=\frac{1}{2}\int_{-1}^{1}\exp\Bigl(\jmath\pi\bigl(\alpha x^2-\beta x\bigr)\Bigr)dx,
\end{equation}
By Completing the square, we have
\begin{equation}
I=\frac{1}{2}\,e^{-\jmath\pi\frac{\beta^2}{4\alpha}}\int_{-1}^{1}\exp\Bigl(\jmath\pi\alpha\Bigl(x-\frac{\beta}{2\alpha}\Bigr)^2\Bigr)dx.
\end{equation}
Let 
\begin{equation}
v=\sqrt{\jmath\pi\alpha}\Bigl(x-\frac{\beta}{2\alpha}\Bigr),
\end{equation}
The integration limits become
\begin{equation}
v_1=\sqrt{\jmath\pi\alpha}\left(-1-\frac{\beta}{2\alpha}\right),\quad
v_2=\sqrt{\jmath\pi\alpha}\left(1-\frac{\beta}{2\alpha}\right).
\end{equation}
Thus,
\begin{equation}
I=\frac{1}{2}\,e^{-\jmath\pi\frac{\beta^2}{4\alpha}}\frac{1}{\sqrt{\jmath\pi\alpha}}
\int_{v_1}^{v_2}\exp(v^2)dv.
\end{equation}
Using the known formula
\begin{equation}
\int \exp(v^2)dv = \frac{\sqrt{\pi}}{2}\operatorname{erfi}(v),
\end{equation}
with \(\operatorname{erfi}(v)=-\jmath\,\operatorname{erf}(\jmath v)\), we obtain the final result
\begin{equation}
\begin{split}
&I=\\
&\frac{e^{\jmath\left(\frac{-\beta^{2}+\alpha}{4 \alpha}\right)\pi}\Biggl[\operatorname{erf}\Bigl(\frac{e^{\jmath\frac{3\pi}{4}}\sqrt{\pi}(\beta-2\alpha)}{2\sqrt{\alpha}}\Bigr)
-\operatorname{erf}\Bigl(\frac{e^{\jmath\frac{3\pi}{4}}\sqrt{\pi}(\beta+2\alpha)}{2\sqrt{\alpha}}\Bigr)
\Biggr]}{4\sqrt{\alpha}}
.
\end{split}
\end{equation}
This completes the derivation.}
\vspace*{3mm}
{\printbibliography}

@ARTICLE{10496996,
  author={Lu, Haiquan and Zeng, Yong and You, Changsheng and Han, Yu and Zhang, Jiayi and Wang, Zhe and Dong, Zhenjun and Jin, Shi and Wang, Cheng-Xiang and Jiang, Tao and You, Xiaohu and Zhang, Rui},
  journal={IEEE Communications Surveys \& Tutorials}, 
  title={A Tutorial on Near-Field XL-MIMO Communications Toward 6G}, 
  year={2024},
  volume={26},
  number={4},
  pages={2213-2257},
  doi={10.1109/COMST.2024.3387749}}

@ARTICLE{10220205,
  author={Liu, Yuanwei and Wang, Zhaolin and Xu, Jiaqi and Ouyang, Chongjun and Mu, Xidong and Schober, Robert},
  journal={IEEE Open Journal of the Communications Society}, 
  title={Near-Field Communications: A Tutorial Review}, 
  year={2023},
  volume={4},
  number={},
  pages={1999-2049},
  doi={10.1109/OJCOMS.2023.3305583}}

@ARTICLE{9693928,
  author={Cui, Mingyao and Dai, Linglong},
  journal={IEEE Transactions on Communications}, 
  title={Channel Estimation for Extremely Large-Scale MIMO: Far-Field or Near-Field?}, 
  year={2022},
  volume={70},
  number={4},
  pages={2663-2677},
  doi={10.1109/TCOMM.2022.3146400}}

@ARTICLE{9738442,
  author={Zhang, Haiyang and Shlezinger, Nir and Guidi, Francesco and Dardari, Davide and Imani, Mohammadreza F. and Eldar, Yonina C.},
  journal={IEEE Transactions on Wireless Communications}, 
  title={Beam Focusing for Near-Field Multiuser MIMO Communications}, 
  year={2022},
  volume={21},
  number={9},
  pages={7476-7490},
  doi={10.1109/TWC.2022.3158894}}

@ARTICLE{10239282,
  author={Wu, Chenyu and You, Changsheng and Liu, Yuanwei and Chen, Li and Shi, Shuo},
  journal={IEEE Transactions on Vehicular Technology}, 
  title={Two-Stage Hierarchical Beam Training for Near-Field Communications}, 
  year={2024},
  volume={73},
  number={2},
  pages={2032-2044},
  doi={10.1109/TVT.2023.3311868}}

@ARTICLE{9913211,
  author={Zhang, Yunpu and Wu, Xun and You, Changsheng},
  journal={IEEE Wireless Communications Letters}, 
  title={Fast Near-Field Beam Training for Extremely Large-Scale Array}, 
  year={2022},
  volume={11},
  number={12},
  pages={2625-2629},
  doi={10.1109/LWC.2022.3212344}}

@ARTICLE{10500334,
  author={Wu, Xun and You, Changsheng and Li, Jiapeng and Zhang, Yunpu},
  journal={IEEE Transactions on Wireless Communications}, 
  title={Near-Field Beam Training: Joint Angle and Range Estimation With DFT Codebook}, 
  year={2024},
  volume={23},
  number={9},
  pages={11890-11903},
  doi={10.1109/TWC.2024.3385749}}

@ARTICLE{7942128,
  author={Selvan, Krishnasamy T. and Janaswamy, Ramakrishna},
  journal={IEEE Antennas and Propagation Magazine}, 
  title={Fraunhofer and Fresnel Distances: Unified derivation for aperture antennas}, 
  year={2017},
  volume={59},
  number={4},
  pages={12-15},
  doi={10.1109/MAP.2017.2706648}}

@ARTICLE{10185619,
  author={Zhang, Yunpu and You, Changsheng and Chen, Li and Zheng, Beixiong},
  journal={IEEE Communications Letters}, 
  title={Mixed Near- and Far-Field Communications for Extremely Large-Scale Array: An Interference Perspective}, 
  year={2023},
  volume={27},
  number={9},
  pages={2496-2500},
  doi={10.1109/LCOMM.2023.3296409}}

@ARTICLE{9112745,
  author={Sarieddeen, Hadi and Saeed, Nasir and Al-Naffouri, Tareq Y. and Alouini, Mohamed-Slim},
  journal={IEEE Communications Magazine}, 
  title={Next Generation Terahertz Communications: A Rendezvous of Sensing, Imaging, and Localization}, 
  year={2020},
  volume={58},
  number={5},
  pages={69-75},
  doi={10.1109/MCOM.001.1900698}}

@ARTICLE{DL_for_beam_training,
  author={Liu, Wang and Ren, Hong and Pan, Cunhua and Wang, Jiangzhou},
  journal={IEEE Communications Letters}, 
  title={Deep Learning Based Beam Training for Extremely Large-Scale Massive MIMO in Near-Field Domain}, 
  year={2023},
  volume={27},
  number={1},
  pages={170-174},
  doi={10.1109/LCOMM.2022.3210042}}

@ARTICLE{10438977,
  author={Jornet, Josep M. and Yang, Nan and Nichols, Roger and Nie, Shuai and Huang, Chongwen and Mallik, Ranjan K.},
  journal={IEEE Wireless Communications}, 
  title={Terahertz Communications and Sensing for 6G and Beyond: How Far Are We?}, 
  year={2024},
  volume={31},
  number={1},
  pages={8-9},
  doi={10.1109/MWC.2024.10438977}}

@ARTICLE{9766110,
  author={Akyildiz, Ian F. and Han, Chong and Hu, Zhifeng and Nie, Shuai and Jornet, Josep Miquel},
  journal={IEEE Transactions on Communications}, 
  title={Terahertz Band Communication: An Old Problem Revisited and Research Directions for the Next Decade}, 
  year={2022},
  volume={70},
  number={6},
  pages={4250-4285},
  doi={10.1109/TCOMM.2022.3171800}}

\end{document}